
\magnification=1200
\hsize 15true cm \hoffset=0.5true cm
\vsize 23true cm
\baselineskip=15pt

\outer\def\beginsection#1\par{\medbreak\bigskip
      \message{#1}\leftline{\bf#1}\nobreak\medskip\vskip-\parskip
      \noindent}

\def \pa {\partial}

\def \La {\Lambda}

\def \a {\alpha}

\def \ga {\gamma}

\def \da {\delta}

\def \r {\rho}

\def \Om {\Omega}
\def \noi {\noindent}

\def\sqr#1#2{{\vcenter{\hrule height.#2pt\hbox{\vrule width.#2pt
height#1pt \kern#1pt\vrule width.#2pt}\hrule height.#2pt}}}

\def\lsim{\mathrel{\rlap{\lower4pt\hbox{\hskip1pt$\sim$}}
    \raise1pt\hbox{$<$}}}         
\def\gsim{\mathrel{\rlap{\lower4pt\hbox{\hskip1pt$\sim$}}
    \raise1pt\hbox{$>$}}}         

 \nopagenumbers

\null\vskip-.5cm
{\hfill\ CERN-TH.6634/92}\par
\line{\hfil DFTT-50/92}
{\hfill\ August 1992}
\vskip2cm
\centerline{\bf BOOSTING AWAY SINGULARITIES FROM}
\centerline{\bf CONFORMAL STRING BACKGROUNDS}
\vskip.5cm
\centerline{\bf M. Gasperini}
\centerline{\it Dipartimento di Fisica Teorica}
\centerline{\it and}
 \centerline{\it INFN, Sezione di Torino, Turin, Italy}
\vskip.5cm
\centerline{\bf J. Maharana}
\centerline{\it Institute of  Physics, Bhubaneswar, 751005,
India}
\vskip.5cm
\centerline{\bf  G. Veneziano }
\centerline{\it Theory Division, CERN, Geneva, Switzerland}
\vskip2cm

\centerline{\bf Abstract}
\noi
Generalizing our previous work, we show how $O(d,d)$ transformations
can be used to "boost away" in new dimensions the physical singularities
that occur generically in  cosmological
and/or black-hole conformal string backgrounds. As an
example, we show how a recent model by Nappi and Witten can be made
singularity-free via $O(3,3)$ boosts involving a fifth
dimension.

\vskip1.5cm

 \noindent
{CERN-TH.6634/92}\par
\noindent
{August 1992}\par
\vfill\eject
\bigskip
\eject

\footline={\hss\rm\folio\hss}
\pageno=1

{\bf 1. Introduction}

A short while ago we have shown [1] how  $O(d,d)$ transformations [2,3]
  acting on anisotropic but
homogeneous (i.e. space-independent) string
cosmologies in $D=d+1$ space-time dimensions, can  turn trivial
 (i.e. flat)
string backgrounds into non-trivial (i.e. curved) ones.
We also noticed [1] that the $O(d,d)$ "boosted" backgrounds came out
 free of curvature singularities for non-exceptional values
 of the boost parameter $\gamma $.

In this note, we generalize the latter observation by showing that,
even if the starting point is a generic, singular cosmology in $D=1+1$
dimensions or a $D=1+1$ black hole, the singularity gets
"boosted away" by $O(2,2)$ transformations
 involving a third, originally flat  dimension.

  We shall then  combine cosmological and black-hole solutions
by considering the inhomogeneous, $D=4$ cosmological model
recently discussed by Nappi and Witten [4], and by  showing that the
singularities of the model can be  boosted
away by $O(3,3)$ transformations
involving a fifth dimension.

\vskip 0.5cm

{\bf 2. Boosting away singularities in $D=2$ backgrounds}

\vskip 0.5cm

The existence of torsion-free ($B=0$) $D=2$ string cosmologies
or black holes is by now well known [5,6]. In a class
of such models  (and in a
convenient reference frame) the corresponding
metric $G_{\mu\nu}$ and dilaton $\phi$ backgrounds are given by:
$$
ds^2 = dx^{\mu}dx^{\nu} G_{\mu \nu} = - dt^2  +
 \tanh^{\pm 2}(\sqrt{\Lambda}~ t/2) ~ d x^2  \; ,\;
\Phi = -\ln \sinh (\sqrt{\Lambda}~ t)
\eqno(1a)
$$
$$
ds^2 =  - \tan^{\pm 2}(\sqrt{\Lambda}~ x/2)~ dt^2  +   d x^2  \; ,\;
\Phi = -\ln \sin (\sqrt{\Lambda}~ x)
 \eqno(1b)
$$
They are exact solutions of the field equations obtained from the low
energy string effective action
$$
S=\int d^Dx \sqrt{|G|} e^{-\phi}(-\La + R + G^{\mu\nu}\pa _\mu \phi
\pa _\nu \phi -{1\over 12}H_{\mu\nu\r}H^{\mu\nu\r})
$$
Here $H_{\mu\nu\r}=\pa_\mu B_{\nu\r}+cyclic$,
$\Phi = \phi - 1/2 \ln \det G$ is the "shifted dilaton", which is
 inert under $O(d,d)$, and $\Lambda$ is the tree-level
cosmological constant:
$$
\Lambda = { (c-c_{crit})\over 3 \alpha '} \;,\;\eqno(2)
$$
where $c_{crit} = 26$ (or $10$) and $c$ is the total central charge
 of the matter fields, $c= 2 + c_{int}$. Of course, when $\Lambda < 0$,
the replacements
$\tan(\sqrt{\Lambda}~ t/2) \leftrightarrow \tanh(\sqrt{-\Lambda}~ t/2)$
 etc. have to be applied
to eqs. (1). The $\pm$ ambiguity occurring in (1)
corresponds to cosmologies (or black holes) related to
each other by scale-factor-duality [7] (or by its equivalent for
black holes), a discrete subgroup of $O(d,d)$. Finally, eq. (1$b$) can
be replaced by a Euclidean black hole simply by replacing $-dt^2$  by
  $ d \tau^2$.

The scalar curvatures corresponding to the geometries (1), (2) are
readily computed to be:

$$
R = \Lambda  (\tanh^{\pm2}(\sqrt{\Lambda}~ t/2) -1) \;\eqno(3a)
$$
$$
R = - \Lambda (\tan^{\pm2}(\sqrt{\Lambda}~ x/2) + 1) \;\eqno(3b)
$$
and thus exhibit singularities at particular values of $t$ or $x$.

Our strategy for removing the singularity (while remaining with a
conformal background) consists in adding
to the spacetime manifold a second flat spatial direction
\footnote{*}{Either this dimension was already present or we have
to readjust the value of $c_{int}$ in order not to change the
value of $\Lambda$.} $z$.
 Since, in both cases, the resulting backgrounds do not depend
  on two (out of the three) coordinates, there will be an $O(2,2)$
group [2,3] acting on the space of such conformal theories.

As already discussed in [1], the space of (gauge-inequivalent) solutions
will be given by the coset $O(2,2)/GL(2)\times B_s$, where $B_s$ is
 the group
of constant shifts of $B$. This coset is described by a single
 parameter $\gamma$, the boost parameter in the plane
spanned by the two above-mentioned coordinates. Under $O(2,2)$:
$$
M \rightarrow \Omega ^T M \Omega   , \;\;~~~~~~~~
 \Phi \rightarrow \Phi ,  \; \eqno(4)
$$
where, as usual [8, 2], $M$ is the $2d$ by $2d$ matrix
$$
M= \pmatrix{G^{-1} & -G^{-1}B \cr
BG^{-1} & G-BG^{-1}B \cr},  \; \eqno(5)
$$
(hereafter $G$ and $B$ stand for the $d$ by $d$ part of the metric and
of the torsion)
and, as in [1], the boost matrix $\Omega$ is taken as
$$
 \Omega (\gamma)  = {1\over 2} \pmatrix { 1 + c & s & c - 1 & - s
\cr
 - s & 1 - c & - s &
 1 + c \cr c - 1 & s & 1 + c & - s \cr s & 1 +c & s & 1 - c \cr },
\; \eqno(6)
 $$
with
$c \equiv \cosh\gamma ,\,\,\, s \equiv
\sinh\gamma ,\,\, \; 0 < \gamma < \infty $.

A straightforward calculation shows that, under the transformation
 (4), the backgrounds  (1$\a$), acquire  non-trivial components in the
$x$--$z$ plane (including a non-vanishing $B$), given by:
$$ G_{\pm}(\gamma ) = \pmatrix { { ( c - 1 ) +
 ( c + 1 ) a^{\pm2}\over { ( c + 1 ) + ( c - 1)
a^{\pm2} }} & {-s ( 1 +a^{\pm2} )\over { ( c + 1) +
  ( c - 1 ) a^{\pm2}}}\cr {- s
 ( 1 + a^{\pm2} )\over { ( c + 1 ) + ( c - 1) a^{\pm2} }} & 1 \cr }$$

$$ B_{\pm}(\gamma ) = \pmatrix { 0 & {- s ( 1 +a^{\pm2} )\over
 { ( c + 1 ) +
 ( c -1 ) a^{\pm2}}}
 \cr {  s ( 1 + a^{\pm2} )\over { ( c + 1 ) + (c - 1 )a^{\pm2} }}
  & 0 \cr }, $$
 $$ \phi_{\pm} (\gamma ) =   -\ln [1+ a^{\pm2}~ \tanh^2 (\gamma /2)]  ~
+  ~ \phi_{\pm} (0)~+~ const \; , \;
 \eqno(7) $$
 where, for the metric ($1a$),
$$
a = a(t) = \tanh (\sqrt{\Lambda}~ t/2) \;
$$
$$
\phi_+(0)=-2 \ln \cosh (\sqrt\La~t/2)~~~,~~~\phi_-(0)=-2 \ln
\sinh (\sqrt\La~t/2).  \eqno(8a)
$$
For the black-hole metric ($1b$) the same result holds, this time in the
$t$--$z$ (or $\tau$--$z$ in the Euclidean case) plane and with:
$$
a = a(x) = \tan(\sqrt{\Lambda}~ x/2) \;
$$
$$
\phi_+(0)=-2 \ln \cos (\sqrt\La~x/2)~~~,~~~\phi_-(0)=-2 \ln
\sin (\sqrt\La~x/2).   \eqno(8b)
$$
 It is straightforward to compute the various curvature tensors
 for the boosted
geometries, choosing e.g. the plus signs.
 For the curvature scalars we find, respectively:
$$
R= {\Lambda\over 2}{ 3 + 4c -7 c^2 -8c \cosh^2(\sqrt{\Lambda}~ t/2)
\over [(c+1)\cosh^2 (\sqrt{\Lambda}~ t/2) +
 (c-1) \sinh^2(\sqrt{\Lambda}~ t/2)]^2}
 \; \eqno (9a)
$$
$$
R= {\Lambda\over 2}{ 3 c^2 + 4c -7  -8c \cos^2(\sqrt{\Lambda}~ x/2)
\over [(c+1)\cos^2 (\sqrt{\Lambda}~ x/2) +
 (c-1) \sin^2(\sqrt{\Lambda}~ x/2)]^2}
 \; \eqno (9b)
$$

Similar expressions hold for the minus-sign choice in eq. (7). We note
 that,
almost magically, all  physical singularities have disappeared from the
dilaton field (7) and from the
scalar curvatures of the boosted metrics for generic values
of the boost parameter $\ga$. We have checked on the computer
that the same is true for the
various components of the Riemann and Ricci tensors, as well as for the
other curvature invariants.

We may ask about the generality of this result,
 in particular whether or not it extends to higher-dimensional
backgrounds. Some explicit examples indicate that, in more general cases,
  singularities are not removed by simply
boosting in an extra dimension.
An example is the one of an isotropic cosmology in $D = d+1$ with $\La=0$
and [5]
$$
   a(t)= (t/t_0)^{-{1\over \sqrt d}}~~~,~~~ \Phi= -\ln(t/t_0)
$$
($t_0$ is an integration constant) which is singular at $t=0$. By
introducing an additional
 flat spatial direction, and by performing the same
transformation as before, one obtains
$$
\phi(\ga)= -(1 + \sqrt d  -{2\over \sqrt d })\ln(t/t_0) -
\ln [(t/t_0)^{{2\over \sqrt d }}+\tanh^2(\ga/2)]
$$
which is still singular at the origin for $d>1$.

On the contrary, our strategy can be applied,
 almost without modifications,
to the four-dimensional model of Nappi and Witten (NW) [4]. This is not
surprising since, as we shall explain below, the NW background itself
can be obtained by an $O(2,2)$ boost
of the direct product of a pair of
two-dimensional models.

\vskip 0.5cm

{\bf 3. $O(2,2)$ derivation and $O(3,3)$ regularization of
the Nappi-Witten

model}

\vskip 0.5cm

Consider a string theory with
vanishing $\Lambda$ [i.e. with $c=c_{crit}$,
see eq. (2)] and containing, besides other degrees of freedom, a
four-dimensional  subspace of Minkowskian signature. A particular
 conformal background
for such a theory consists of the direct product of a cosmological
metric and of a Euclidean black hole, each one living
 in a two-dimensional subspace. The line element of such a model
is thus a particular combination of metrics of the type
(1$a$) and (1$b$):
$$
ds^2 = dx^{\mu}dx^{\nu} G_{\mu \nu} = - dt^2  +  d x^2  +
\tan^{-2}(\sqrt{V}~ t/2) ~ d y^2  +
\tan^{ 2}(\sqrt{V}~ x/2)~ dz^2  ,
$$
$$
\Phi = -\ln \sin(\sqrt{V}~ t) -\ln \sin(\sqrt{V}~ x).
 \;\eqno(10)
$$
where $V$ is a positive constant.

Taken by itself, the black-hole metric contained in (10) would
require a positive
cosmological constant $\Lambda=V$, while the cosmological-type
background would need an opposite value for $\Lambda$.
The complete background (10) is thus conformal for $\Lambda = 0$.
Obviously, many other possibilities, with or without an overall
cosmological constant, can be considered. For the sake of
definiteness and in order to
make contact with NW, we shall take in the
 following $V = 4$, in usual
string units $2 \alpha ' =1$.

Since the metric in (10) is independent of $y, z$, we
 can apply to it any $O(2,2)$ transformation to get new solutions.
After moding out by gauge transformations we are left, as before,
with a one-parameter family of gauge-inequivalent backgrounds given by:
$$
M(\delta) =  \Omega ^T(\delta)  M \Omega(\delta),  \;\eqno(11)
$$
where:
$$
\Omega (\delta) =
      {1\over \sqrt 2} \pmatrix { \delta &  0 & 0 & \delta
\cr
 0 & 1  & - 1 &
 0 \cr 0 & \delta ^{-1} & \delta ^{-1} & 0 \cr -1 & 0 & 0 & 1  \cr }
\;\; \eqno(12)
$$
is an $O(2,2)$ matrix. It is easy to check that $M(\da)$ reproduces
precisely \footnote{*}
{This observation was first made by A. Giveon,
 see Note Added in ref. [4]. }
 the family of inhomogeneous NW cosmologies after the following
  identification of their parameter $\alpha$:
$$
\delta ^2 = { 1 -\sin \alpha \over 1+ \sin \alpha}   \; \eqno(13)
$$

We see here a good example of the power of $O(d,d)$
 in generating highly non-trivial conformal backgrounds. Nonetheless, if
no extra dimension is called in, the boosts fail to remove
 the singularities
  occurring in the original background. As noticed by NW, their
cosmological solutions do exhibit curvature singularities, for instance
at
$t=0= x$. Note, however, that the original background
(10) was even more singular than the one of NW, since it had curvature
singularities even at finite $x$ for $t \rightarrow 0$ (and similar ones
at special values of $x$ and generic $t$).

We shall now show how to altogether eliminate  the singularities
 of the NW background [or of that of eq. (10)] by introducing a fourth,
 originally flat spatial direction, parametrized by the
coordinate $w$. The theory now acquires a larger symmetry
 isomorphic to $O(3,3)$.

Given the results of Section 2, it is natural to try to boost away the
singularities of the NW model by performing, successively, a boost
in the $z$--$w$ plane and one in the $y$--$w$ plane.
The first will certainly remove the singularities along the $x$ axis,
simply by repeating the steps of Section 2.
 The nice surprise is that the second boost eliminates also the
singularities along the time direction
without introducing back the ones already removed.

We shall now give some details of the actual calculations. Starting from
  the matrix $M$ corresponding to the metric (10) and $B=0$,
the boosted backgrounds are   given in terms of a boosted $M$   by:
$$
\tilde M (\gamma _1, \gamma _2) = \Omega ^T(\gamma _2)
 \Omega ^T(\gamma _1)
 M \Omega (\gamma _1) \Omega  (\gamma _2), \;\eqno(14)
$$
where $\Omega (\gamma _1)$ is the $z$--$w$ boost [here, in analogy with
eq. (6),  we use the notation $c_1 \equiv \cosh \gamma_1$ etc.]:
$$
\Omega (\gamma _1) =  {1\over 2} \pmatrix { 2 & 0 & 0 & 0 & 0 & 0 \cr
0 & 1 + c_1 & s_1 & 0 & c_1 - 1 & - s_1
\cr  0 &
 - s_1 & 1 - c_1 & 0 & - s_1 &
 1 + c_1 \cr 0 & 0 & 0 & 2 & 0 & 0 \cr 0 & c_1 - 1 & s_1 & 0 & 1 +
  c_1 & - s_1
\cr  0 & s_1 & 1 +c_1 & 0 & s_1 & 1 - c_1 \cr },
\; \eqno(15)
$$
 and $\Omega (\gamma _2)$ is a similar $y$--$w$ boost:
$$
\Omega (\gamma _2) = {1\over 2} \pmatrix {
 1 + c_2 & 0  & s_2 &  c_2 - 1 & 0  & - s_2 \cr 0 & 2 & 0 & 0 & 0 & 0 \cr
 - s_2 & 0 & 1 - c_2  & - s_2 & 0 &
 1 + c_2 \cr  c_2 - 1 & 0 & s_2 &  1 + c_2 & 0 & - s_2
\cr  0 & 0 & 0 & 0 & 2 & 0 \cr  s_2 & 0 & 1 +c_2 &  s_2 & 0 &
 1 - c_2 \cr }.
\; \eqno(16)
$$

Computations can be done easily with some computer help and lead
to the following doubly-boosted backgrounds (for simplicity $\ga_1=\ga_2
=\ga$, which is already sufficient for our purpose)
$$
\tilde G_{11} (\gamma ) =  {c-1+(c+1)a^2 \over c+1+(c-1)a^2}~~~~,
{}~~~~\tilde G_{12}(\ga)= {s^2(a^2-1)(b^2+1)\over
[c+1+(c-1)a^2][c+1+(c-1)b^2]}
$$
$$
\tilde G_{13}(\ga)= -{s(1+a^2)\over c+1+(c-1)a^2}~~~~,
{}~~~~\tilde G_{22}(\ga)= {c-1+(c+1)b^2\over c+1+(c-1)b^2}
$$
$$
\tilde G_{23}(\ga)=
-{s(1+b^2)[c(a^2-1)-a^2-1]\over [c+1+(c-1)a^2][c+1+(c-1)b^2]}~~~~,
{}~~~~ \tilde G_{33}(\ga)=1
$$
\vskip 1 true cm
$$
\tilde B (\gamma ) = \pmatrix{ 0
& -\tilde G_{12}(\ga)
& \tilde G_{13}(\ga) \cr
\tilde G_{12}(\ga)
& 0
& \tilde G_{23}(\ga) \cr
-\tilde G_{13}(\ga)
& -\tilde G_{23}(\ga)
& 0 \cr }
$$
\vskip 1 true cm
$$
\tilde \phi (\gamma) =  \Phi + \ln {4ab \over
[c+1+(c-1)a^2][c+1+(c-1)b^2]}
\eqno(17)
$$
(the indices $1,2,3$ run over the set of coordinates $(y,z,w)$).
Here $a^2=\tan^{-2}(t)$, $b^2=\tan^2(x)$, and $\Phi$ is the shifted
dilaton of eq.(10), but the result (17) holds generally for any
background whose metric may be written in the form $G_{\mu\nu}=
diag~ (-1,1,a^2,b^2)$.

At this point various components of the
 curvature tensor and their contractions can be
computed with the help of a  program (we have used MACSYMA) and
some have been double checked either analytically or through consistency
with the equations of motion (vanishing $\beta$-function conditions).
In all cases we have found no singularity occurring in the boosted
metrics.

As an example we quote here the curvature scalar and the coupling
constant, which are given respectively by:
$$
R = {16 \cos^2t \sin^2t -20 s^2 \over [(c+1)\sin^2t +(c-1)\cos^2 t]^2}
+{20 s^2 -16 \cos^2x \sin^2x \over [(c+1)\cos^2x +(c-1)\sin^2x]^2}
$$
$$
e^{\phi} = [(c+1)\sin^2t +(c-1)\cos^2 t]^{-1} [(c+1)\cos^2x +
(c-1)\sin^2x]^
{-1}. \;
\eqno(18)
$$

Strictly speaking, in order to obtain a NW-like model, we
should still perform a third boost in the $y$--$z$ plane, with the same
$\Om (\da)$ as in eq. (12). This would
complicate the result (17), but it is quite obvious that
it would not alter the conclusion that all singularities are
indeed removed.

This result is not in contradiction, of course, with the classical
singularity theorems [9]. By computing the components of the Ricci
tensor for the doubly-boosted metric we have indeed, from eqs. (17),
$$
\tilde R_{00}= {2s^2+4(2c~\cos^2t-c-1) \over
[(c+1)\sin^2t+(c-1)\cos^2t]^2}. \;
\eqno(19)
$$
If $c=1$, the expression $R_{00}=-2/\sin^2t$ is recovered. It is
  valid for the
original torsionless background of eq. (10). In that case $R_{00}$ is
always negative, so that the strong energy condition is everywhere
satisfied. If, instead, $c>1$, one may see from eq. (19) that the
sign of $R_{00}$ changes around the points that correspond to
singularities of the original metric (such as, for instance, $\cos
t=1$).
The strong-energy condition is thus violated, and this explains why the
singularity theorems can be evaded. Moreover, the presence of torsion in
the boosted background (17) seems to stress the crucial role played by
this field in avoiding singularities, in agreement with
 our previous conjecture [1].
\vskip 1true cm
In conclusion, we have seen: i) how $O(d,d)$
transformations acting within the space-time dimensions of a given
singular background can generate new non-trivial conformal
backgrounds, without major effects on the singularities themselves,
and ii) how $O(d+1,d+1)$ transformations involving an extra dimension
(with originally trivial geometry) can instead "boost away" the
original singularities by smearing out the curvature
 over a range of the original
variables, and by involving in a non-trivial way the new dimension.

The full meaning and generality of our results remain to be understood.
It would  be nice, for instance,
 to know if they depend heavily on having
started with a direct product of $D=2$ backgrounds.
We have seen that the most naive generalization of our procedure
to genuine $D>2$ backgrounds does not work, but we cannot exclude
that a more general technique will be able to
 eliminate the singularities.
 In any event, we
believe  the phenomenon we have described to be a positive step in the
quest for singularity-free and more realistic cosmological
string scenarios.
 \vskip 1 true cm
One of us (J.M.) would like to thank the TH division at CERN
for its warm hospitality while part of this work was done.

\vfill\eject

\centerline{\bf References}
\vskip 1 cm
\item{1.}M. Gasperini, J. Maharana and G. Veneziano, Phys. Lett.
   B 272 (1991) 277.

\item{2.}K.A. Meissner and G. Veneziano, Phys. Lett. B267 (1991) 33;
  Mod. Phys. Lett. A6 (1991) 3397;

 M. Gasperini and G. Veneziano, Phys. Lett. B 277 (1992) 256.

\item{3.}A. Sen, Phys. Lett. B271 (1991) 295;

S.F. Hassan and A. Sen, Nucl. Phys. B375 (1992) 103;

 A. Giveon and M. Rocek, Nucl. Phys. B380 (1992) 128;

J. Horne, G. Horowitz and A. Steif, Phys. Rev. Lett. 68 (1992) 568;

J. Panvel, Phys. Lett. B284 (1992) 50;

J. Maharana and J. H. Schwarz, "Non-compact symmetries in string theory",

Caltech preprint CALT-68-1790 (1992);

A. Kumar, "Gauged string actions and $O(d,d)$ transformations", preprint

CERN/TH.6530/92 (1992);

S. Kar, S. Pratik Khastgir and S. Sengupta, "Four dimensional stringy

black membrane", Bhubaneswar preprint IP/BBSR/92-35;

S. Mohapatra, "Four dimensional 2-brane solution in chirally

gauged Wess-Zumino-Witten models", Tata preprint, TIFR/TH/92-28;

S. Kar, A. Kumar and G. Sengupta, "Hidden isometry and a chiral gauged

WZW model",  Bhubaneswar preprint IP/BBSR/92-67.

\item{4.}C. R. Nappi and E. Witten, "A closed, expanding universe
in string theory", Princeton preprint IASSNS-HEP-92/38, (1992).

\item{5.} M. Mueller, Nucl. Phys. B337 (1990) 37;

  G. Veneziano, Phys. Lett. B265 (1991) 287.

\item{6.} K. Bardakci, A. Forge and E. Rabinovici, Nucl. Phys.
 B344 (1990) 344;

S. Elitzur, A. Forge and E. Rabinovici, Nucl. Phys. B359 (1991) 581;

I. Bars and D. Nemeschansky, Nucl. Phys. B348 (1991) 89;

G. Mandal, A.M. Sengupta and S.R. Wadia, Mod. Phys. Lett. A6 (1991)

1685;

E. Witten, Phys. Rev. D44 (1991) 314.

\item{7.} G. Veneziano, ref. [5];

A.A. Tseytlin, Mod. Phys. Lett. A6 (1991) 1721;

A.A. Tseytlyn, in Proc.  First Int. A.D. Sakharov Conference on Physics,

 ed. by L.V. Keldysh et al. (Nova Science Pub., Commack, NY, 1991);

A.A. Tseytlin and C. Vafa, Nucl. Phys. B372 (1992) 443;

A. Giveon, Mod. Phys. Lett. A6 (1991) 2843;

R. Dijkgraaf, E. Verlinde and H. Verlinde, Nucl. Phys. B371 (1992) 269;

E. Smith and J. Polchinski, Phys. Lett. B263 (1991) 59;

T. Kugo and B. Zwiebach, Prog. Theor. Phys. 87 (1992) 801.

 \item{8.}A. Giveon, E. Rabinovici and G. Veneziano,
  Nucl. Phys. B322 (1989) 167;

A. Shapere  and F. Wilczek, Nucl. Phys. B320 (1989) 669.

\item{9.} S. W. Hawking and G.R.F. Ellis, "The large scale structure of
spacetime", (Cambridge Univ. Press, Cambridge, 1973).

\

 \end